\documentclass{camera}
\usepackage{epsfig}

\begin{document}
\renewcommand{\thefootnote}{\fnsymbol{footnote}}

%
\title{$\sigma( e^+ e^- \to hadrons)$ AT LOW ENERGY: EXPERIMENTAL STATUS
  AND PROSPECTS FOR THE FUTURE. \\ ITS INFLUENCE ON $\alpha_{QED}(M^2_{Z})$ AND
$(g-2)_{\mu}^{\,\,\,^*}$}

%
\author{Graziano Venanzoni}

%
\organization{Universit\`a and Sezione INFN, Pisa, Italy \\
e-mail:Graziano.Venanzoni@pi.infn.it}

\maketitle

%
\begin{abstract}
In this talk I will review the recent experimental status of $\sigma(
e^+ e^- \to hadrons)$ at $\sqrt{s}<10 GeV$ and the prospects for the future.
The influence on $\alpha_{QED}(M^2_{Z})$ and $(g-2)_{\mu}$ is also discussed.
\end{abstract}
\footnotetext[1]{Invited talk given at the XIII italian meeting on
  high energy physics ``LEPTRE'', Rome, April 18-20 2001, to appear in the
  proceedings.}

\section{Why $\sigma(e^+ e^- \to hadrons)$ at low energy is still interesting?}
An undeniable trend of the high energy physics community is the
exploration of high energy ranges by constructing more and more powerful
machines and detectors. However, beside that, there is still a considerable
effort on precise physics at low energies, which uses $e^+e^-$
annihilation in the region below 10 GeV: DA$\Phi$NE and VEPP-2M 
at $\sqrt{s} <1.4 \,GeV$; BEPC at $2<\sqrt{s} <5 \,GeV$
and CSR, KEKB, and PEP-II colliders at $\sqrt{s}=10 \,GeV$. 
Though the main motivation for $\phi$ and $B$ 
factories concerns the CP violation
studies, R-measurement at low energy ($R=\frac{\sigma(e^+ e^- \to hadrons)}{\sigma(e^+ e^-
    \to \mu^+ \mu^-)}$) is renewing its interest due to the precision 
  reached to test the Standard Model at LEP and SLC and also to the
  new experimental result of $(g-2)_{\mu}$ at Brookhaven. The experimental 
accuracy reached so far asks for a precise determination of the theoretical
estimation of both  $\alpha_{QED} (M^2_{Z})$ and $(g-2)_{\mu}$, whose main
error comes from the non-perturbative computation of the hadronic
contributions, which can be computed by using $e^+ e^-$ data at low energy.
A precise measurement of R in this region is therefore mandatory, and
is also one of the main reason for new projects:
VEPP2000 ($\sqrt{s} <2 \,GeV$), PEP-N ($1.4<\sqrt{s} <2.5 \,GeV$), 
BEPCII ($2<\sqrt{s} <5 \,GeV$) , CLEO-C($3<\sqrt{s} <5 \,GeV$).


\begin{table}
\begin{center}
\noindent
\begin{tabular}[h]{|c|c|c|c|c|c|}
\hline\hline
Place & Ring & Detector & $\sqrt{s} (GeV)$ &  pts & Year \\
\hline
Novosibirsk & VEPP-2M & CMD2, SND & 0.28-1.4 & 128 & '97-'99 \\
             & VEPP-2 & OLYA, ND, CMD & 0.28-1.4 & - & '78-'87 \\
\hline
Frascati & Adone & $\gamma\gamma$2, MEA, BOSON, BCF & 1.42-3.09 & 31 &
              '78-'82\\
\hline
Orsay & DCI & M3N,DM1,DM2 & 1.35-2.13 & 33 & 1978 \\
\hline
Beijing &BEPC& BESII & 2-5 & 85 & 1998-99 \\
\hline
SLAC & Spear & MARKI & 2.8-7.8 & 78 & 1982 \\
\hline
Hamburg & DORIS & DASP &3.1-5.2 & 64 & 1979 \\
        &       & PLUTO & 3.6-4.8,9.46 & 27 & 1977 \\
        &       & C.BALL & 5.0-7.4 & 11 & 1990 \\
        &       & LENA & 7.4-9.4 & 95 & 1982 \\
\hline
Novosibirsk & VEPP-4 & MD-1 & 7.23-10.34 & 30 & 1991 \\
\hline
\end{tabular}
\caption{Overview of R measurements.}
\label{tab1}
\end{center}
\end{table}

\begin{figure}[h]
\begin{tabular}{cc}
\epsfig{file=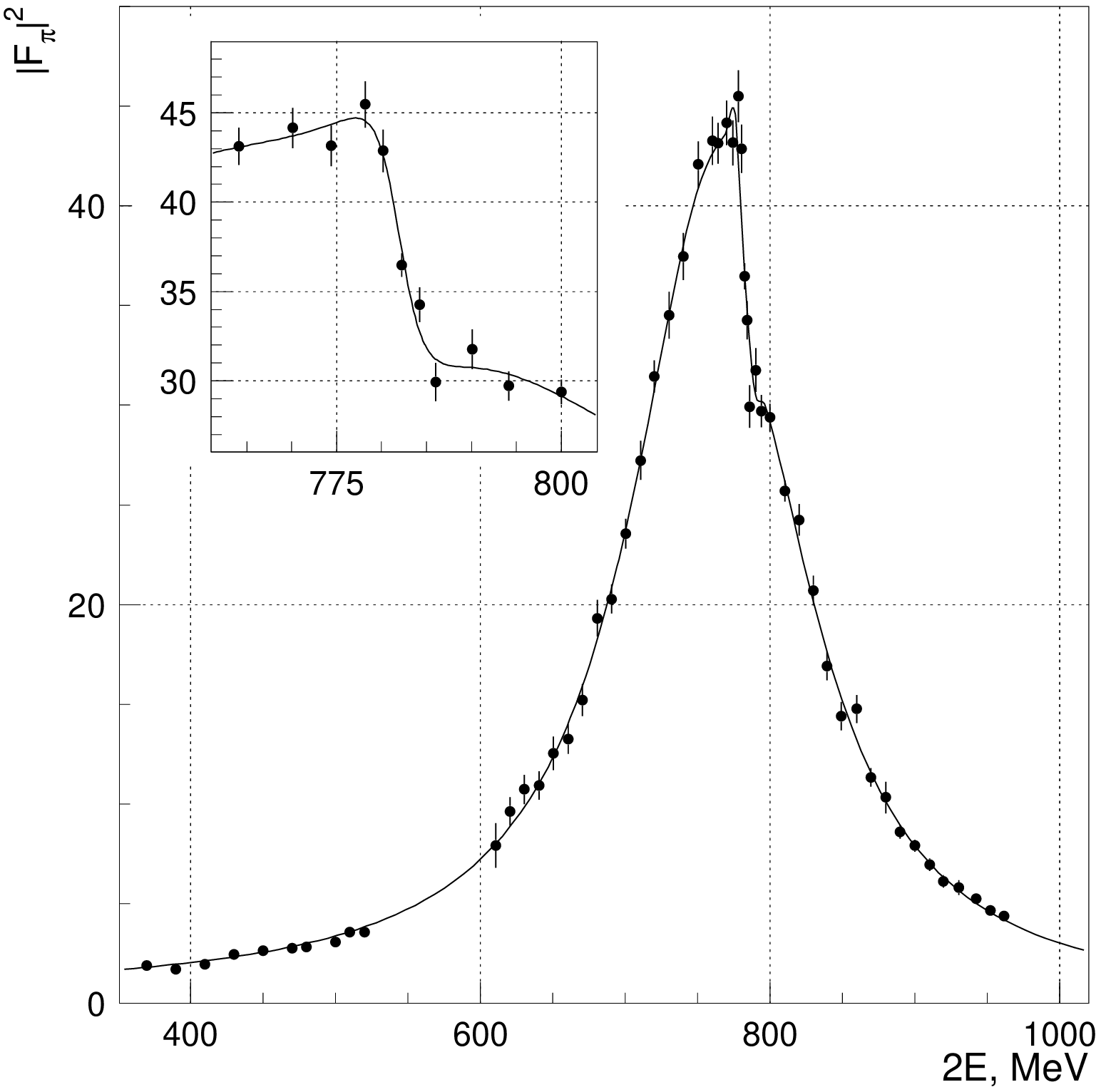,width=8cm,height=8cm} &
\epsfig{file=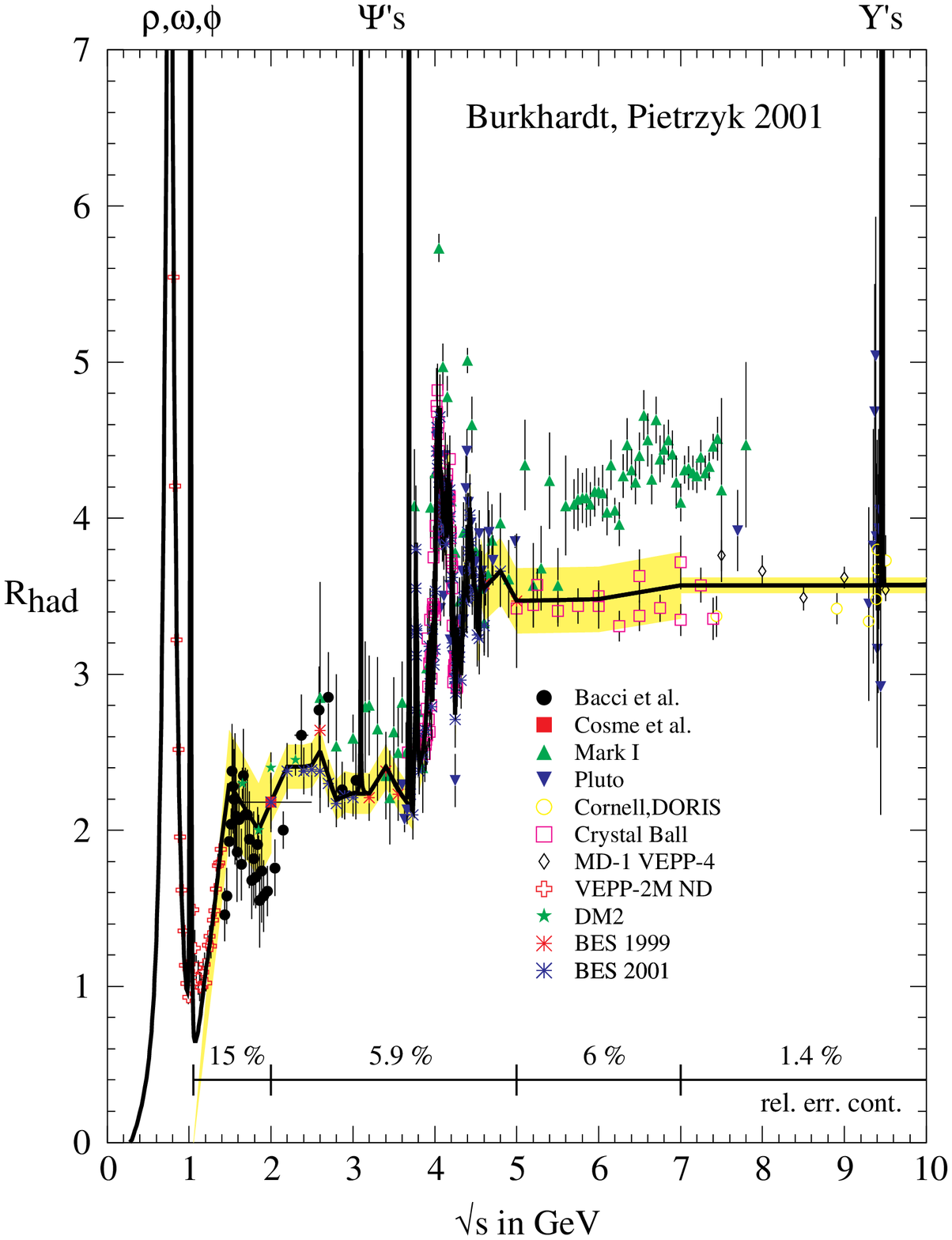,width=8cm,height=8cm} \\
\end{tabular}
\caption{{\it Left}: measurement of the pion form factor from CMD-2. {\it
    Right}: an updated compilation of R-measurement from H. Burkhardt and B. 
Pietrzyk~\cite{bp}.}
\label{fig1}
\end{figure}

R has been measured by many laboratories in the last 20
years, as shown in Tab.~\ref{tab1}. 
Fig.~\ref{fig1}(b) shows
an up-date compilation of these data done by Burkhart and Pietrzyk~\cite{bp}.
The main improvements come in the region below 5 GeV, in particular
between $2-5\,GeV$ where the BESII coll. has reduced the error 
to $\sim 7\%$ (before was $\sim 15\%$), 
and in the region below 1 GeV, where the CMD2 coll. has measured
the pion form factor with a  systematical error of $1.4\%$ 
(see Fig.~\ref{fig1}(a)). Both these new results have  significant impact on 
the updated calculation of $\alpha_{QED} (M^2_{Z})$ and $(g-2)_{\mu}$.
While the data between 2-5 GeV are now closer to perturbative QCD,
the error in the 1-2 GeV region is still  15\%:
a reduction of  this error to few {\it percent} will be very important
both for $\alpha_{QED} (M^2_{Z})$ and $(g-2)_{\mu}$ calculations.
\section{Effective $\alpha_{QED}$ and precision test of the Standard Model}
The precision reached for the measurements performed at LEP and SLC allows 
a stringent test of the Standard Model and to predict the
Higgs mass. 
As discussed many times  in this conference, the QED coupling
constant at $\sqrt{s}=M_{Z}$, $\alpha_{QED}(M^2_{Z})$, is now the
limiting factor for the fit of the SM. The uncertainity of
$\alpha_{QED}(M^2_{Z})$ arises from the low energy contribution of the five
quarks, $\Delta\alpha^{(5)}_{had}(M^2_{Z})$,
 which cannot be reliably calculated using perturbative QCD:
\[
\alpha(M^2_{Z}) =\frac{\alpha(0)}{1-\Delta\alpha_l(M^2_{Z})-
\Delta\alpha^{(5)}_{had}(M^2_{Z})-\Delta\alpha_{top}(M^2_{Z})}
\]
The leptonic contribution is computed to the third order, while 
the top contribution depends on the mass of the top quark, which is a
parameter of the fit.

The hadronic contribution $\Delta\alpha^{(5)}_{had}(M^2_{Z})$ can be however
evaluated by using $e^+ e^-$ data, {\it via} a dispersion integral:
\begin{eqnarray}
\Delta\alpha^{(5)}_{had}(M^2_{Z}) &=& -\frac{\alpha M^2_{Z}}{3\pi} Re 
\int_{4 m^2_{\pi}}^{\infty} ds\frac{R(s)}{s(s-M^2_{Z} -i\epsilon)} = \\
& = &  -\frac{\alpha M^2_{Z}}{3\pi} \Big(
 Re \int_{4 m^2_{\pi}}^{E_{cut}^2} ds\frac{R^{data}(s)}{s(s-M^2_{Z} -i\epsilon)}
+ Re \int_{E_{cut}^2}^{\infty} ds\frac{R^{pQCD}(s)}{s(s-M^2_{Z} -i\epsilon)} 
\Big)
\end{eqnarray}
The above integral has been intentionally split into two parts to
emphasize the role of the 
energy cut above which perturbative QCD (pQCD) is used:
theoretical computation of 
$\Delta\alpha^{(5)}_{had}(M^2_{Z})$ 
 depends not only on
the experimental precision on $R^{data}(s)$, but also on the choice of the
energy cut, leading to different predictions~\cite{fj}.
\section{Hadronic contribution to the anomalous magnetic moment of the muon}
In February 2001, Farley and colleagues~\cite{far}, reported a new
experimental value of the anomalous magnetic moment of the muon
 $a_{\mu}= (g-2)/2 = (11 659 202\pm14\pm6)\times 10^{-10}$ using
a positive muon beam, which is $2.6\sigma$ away from what is expected from
 SM~\cite{mar}. Also in this case the main contribution to the theoretical
 error is given by the low energy hadronic contribution to the 
vacuum polarization, which again can be computed using
experimental $e^+ e^-$ data:
\[
a_{\mu}^{had}=(\frac{\alpha m_{\mu}}{3\pi})^2
 \Big(
  \int_{4 m^2_{\pi}}^{E_{cut}^2} ds\frac{R^{data}(s)\hat K(s)}{s^2}
+  \int_{E_{cut}^2}^{\infty} ds\frac{R^{pQCD}(s)\hat K(s)}{s^2} 
\Big)
\]
The kernel $\hat K(s)$  is a smooth bounded function; the $1/s^2$
dependence in the above integral
enhances low energy contributions, {\it i.e.} mainly 
$\sqrt{s}< 1 \,GeV$ (the $\rho$ contributes for 62\% 
of $a_{\mu}^{had}$). 
Recent evaluations have been computed using different
approaches; a conservative data based approach using new data from BESII
and CMD-2 found $a_{\mu}^{had}=(698.75 \pm 11.11)\times 10^{-10}$~\cite{fj} 
with an error still dominated by the  $\sqrt{s}< 1.4 \,GeV$ region.

\section{Comments on the recents experimental results}
\subsubsection*{Recent results from VEPP-2M}
Many hadronic channel were measured at VEPP-2M by CMD2 and SND
collaborations in the region 0.4-1.4 GeV, as shown  Tab.~\ref{tab2}.
\begin{table}
\begin{center}
\noindent
\begin{tabular}[h]{|c|c|c|c|}
\hline\hline
 Channel & Energy range & Syst. Error & Experiment \\
\hline
$e^+ e^-\to\pi^+\pi^-$ & 0.61-0.96 & 1.4\% & CMD-2 \\
\hline
$e^+ e^-\to\pi^+\pi^-\pi^0$ & 0.76-0.81 & 1.3\% & CMD-2 \\
\hline
$e^+ e^-\to\pi^+\pi^-\pi^0$ & 0.98-1.06 & 3-5\%  &
 SND, CMD2 \\
\hline
$e^+ e^-\to\pi^+\pi^-\pi^0$ & 1.04-1.38 & ~12\% &
 SND \\
\hline
$e^+ e^-\to\pi^+\pi^-\pi^+\pi^-$ & 0.67-0.97 & ~12\% &
 CMD-2 \\
\hline
$e^+ e^-\to 2\pi^+ 2\pi^-,\pi^+\pi^- 2\pi^0$& 1.05-1.38 & ~15\% &
 CMD-2, SND \\
\hline
$e^+ e^-\to\pi^+\pi^-\pi^+\pi^-\pi^0$ & 1.28-1.38 &~15\%  &
 CMD-2  \\
\hline
$e^+ e^-\to K^+ K^-, K_S K_L$& 0.98-1.06 & 
3-5\%  &   SND, CMD2 \\
$e^+ e^-\to K_S K_L$ & 1-1.4 & $\sim 15\%$ & SND \\
\hline\hline
\end{tabular}
\caption{Recent results on multi hadronic channels at VEPP-2M.}
\label{tab2}
\end{center}
\end{table}
As said before, the main contribution for $a_{\mu}$ comes from the
region below 1.4 GeV, in particular from the $e^+ e^-\to\pi^+\pi^-$
channel; 
the systematical error is 1.4\%, and  it's expected to go down to 0.6\%
in the near future.
In order to achieve such a precision the systematics were carefully
checked, for example the error coming from the
 energy beam is reduced by the resonance depolarisation
technique.
The main contribution to the systematic error comes now from the theoretical
uncertainty to the radiative corrections: keeping the error below 1\% is a
challenging task.

\subsubsection*{Recent results from BESII at BEPC}
The BESII collaboration has recently published a new measurement of R in
the region 2-5 GeV, based on 85 points taken between February and June 99,
 with an average precision of 6.6\%, a factor 2 better of the previous
results~\cite{bes}. R was determined inclusively, from the number of observed hadronic
events, $N_{had}^{obs}$:
\[
R=\frac{N_{had}^{obs}-N_{bckg}-\sum_l N_{ll}-N_{\gamma\gamma}}
{\sigma^0_{\mu \mu}\cdot L \cdot \epsilon_{had}\cdot (1+\delta)}
\]
where $N_{bckg}$ is the number of beam-associated background events; 
$\sum_l N_{ll}$  and $N_{\gamma\gamma}$ are respectively 
the background coming from misidentified events in one and two photons
processes; $L$ is the integrated luminosity; $\delta$ is the radiative
correction and $\epsilon_{had}$ is the overall detector efficiency.
In order to keep the error to $\sim 7\%$ a big effort was done 
on: (a) Monte Carlo simulation to better understand detector efficiency;
(b) estimation of $N_{bckg}$ by means of separated beam and single beam
  operation; (c) radiative correction by comparing different schemes.


\section{Conclusion: what we expect from the future?}
We will now conclude by showing what we expect in the next years on 
hadronic cross section measurements at low energy:

\underline {0.4-1.4 GeV region}

\begin{itemize}
\item {\bf DA$\Phi$NE - LNF-Frascati (KLOE)}:

 - Measurement of $|F_{\pi}|^2$ at $\sqrt{s}< 1 \,GeV$ {\it via}
  radiative return~\cite{cat}; 

- upgrade for energy scan (2004?).

\item {\bf VEPP2M -Novosibirsk (CMD2, SND)}:

- Measurement of $|F_{\pi}|^2$ with 0.6\% of systematic error 
and refined results on other channels;
%

- new collider proposed VEPP-2000 (2003?) with $\sqrt{s}$ up to 2 GeV and
  expected luminosity of $10^{31}-10^{32} cm^{-2} sec^{-1}$;
\end{itemize}
\underline {1.4 - 2.5 GeV region}

{\bf PEP-N: new asymmetric $e^+ e^-$ collider proposed at SLAC (2005?)}

(http://www.slac.stanford.edu/grp/rd/epac/LOI/): expected 
$\delta_R/R\sim 2.5\%$; 

%
%
%
%
\vspace{.5cm}
\underline {2 - 5 GeV}:

\begin{itemize}
 \item {\bf BEPC and BES upgraded (BEPCII and BESIII)}: 
 expected $\delta_R/R \leq 3\%$;
%

\item {\bf CLEO-C}: Modify CESR for high L in 3-5 GeV region (2003?)

(http://www.lns.cornell.edu/public/CLEO/CLEO-C/index.html)

\end{itemize}

\underline {below 10 GeV}:

Use ISR at {\bf B-factories} to scan the region below 
$\Upsilon(4s)$~\cite{sol}

\vspace{.3cm}
All the future results (within the existing or with the new detectors)
will contribute to determine an new exciting era
for the hadronic cross section measurement.

\vspace{.5cm}
{\bf Acknowledgement}: I wish to thank 
Fred Jegerlehner, Wolfgang Kluge and 
Achim Denig for very useful discussions.

\end{document}